\documentclass[preprint,superscriptaddress,frontmatterverbose,aps,prd,amsmath,amssymb,floatfix,notitlepage,nofootinbib]{revtex4-2}

\usepackage[normalem]{ulem}
\usepackage{xcolor}
\usepackage{graphicx}
\usepackage{dcolumn}
\usepackage{orcidlink}
\usepackage{cancel}

\allowdisplaybreaks

\def\be{\begin{equation}}
\def\ee{\end{equation}}
\def\bea{\begin{eqnarray}}
\def\eea{\end{eqnarray}}

\def\beq{\begin{equation}}
\def\eeq{\end{equation}}
\def\bea{\begin{eqnarray}}
\def\eea{\end{eqnarray}}

\def\roughly#1{\mathrel{\raise.3ex\hbox
{$#1$\kern-.75em\lower1ex\hbox{$\sim$}}}}

\def\order{\lower 1.8ex \hbox{\LARGE\~{}}}

\begin{document}

\title{Dark Higgs in Hidden Sector as a Probe for Dark Matter}
\author{Faeq Abed\,\orcidlink{0000-0002-3934-4367}}
\email{faeq.abed@irsra.gov.iq}
\affiliation{Particle Physics Lab, Iraqi Nuclear Regulatory Authority, Nidhal, Bghd 28042, Iraq}

\author{Asmaa AlMellah\,\orcidlink{0000-0003-3975-3504}}
\email{asmaa.almellah@gmail.com}
\affiliation{Department of Physics, Faculty of Engineering and Natural Sciences, S\"uleyman Demirel University, Isparta, Turkey 32260}

\author{Gaber Faisel\,\orcidlink{0000-0001-8770-1966}}
\email{gaberfaisel@sdu.edu.tr}
\affiliation{Department of Physics, Faculty of Engineering and Natural Sciences, S\"uleyman Demirel University, Isparta, Turkey 32260}

\begin{abstract}
This study presents a sensitivity analysis of exotic Higgs boson decays at the electron--positron stage of the Future Circular Collider (FCC-ee), performed within the FCCAnalyses framework. The analysis investigates Higgs boson production in association with a Z boson in electron--positron collisions at a center-of-mass energy of 240~GeV. The Higgs boson is assumed to decay into a pair of long-lived scalar particles, while the Z boson decays leptonically. A hadronic final state is considered, in which the long-lived scalars subsequently decay into bottom--antibottom quark pairs.

The simulation chain is implemented using the FCCAnalyses framework, with event generation performed using \textsc{MadGraph} and \textsc{Pythia}, and detector effects modeled with \textsc{Delphes}. Displaced vertices arising from the decays of long-lived particles are reconstructed using the FCCAnalyses implementation of the \textsc{LCFIPlus} secondary vertex finding algorithm, enhanced with extended features such as customized track selection.

The final event selection requires the reconstruction of a Z boson together with at least two displaced vertices. This strategy efficiently suppresses Standard Model backgrounds while retaining at least three expected signal events, including statistical uncertainties, over most of the explored parameter space. The study considers scalar masses of 20~GeV and 60~GeV and mixing angles of $10^{-5}$, $10^{-6}$, and $10^{-7}$. The results demonstrate that FCC-ee will be sensitive to long-lived scalars with decay lengths ranging from approximately 1~mm to 10~m, with optimal sensitivity around a decay length of 0.3~m.

\end{abstract}

\maketitle
\newpage

\section{{ Introduction }}

A substantial body of research has established that all matter in the Universe is composed of a limited set of fundamental particles governed by four fundamental interactions: the electromagnetic, strong, and weak forces, as well as gravity. Three of these forces—excluding gravity—together with all known elementary particles are described by the Standard Model (SM) of particle physics. The SM has achieved remarkable success in accurately predicting a wide range of experimental results. Nevertheless, several key observations, such as the matter--antimatter asymmetry of the Universe and the existence of dark matter, remain unexplained within the SM framework, indicating the necessity for new physics beyond the Standard Model (BSM)~\cite{CERN:StandardModel}.

The Large Hadron Collider (LHC) is currently the world’s most powerful particle accelerator, colliding proton or heavy-ion beams at center-of-mass energies in the multi-TeV range, and is expected to operate until 2026. The collider hosts four main interaction points, where the ATLAS, CMS, ALICE, and LHCb experiments are located to investigate the properties of particles produced in high-energy proton--proton collisions~\cite{CERN:LHC}. To extend and complement the physics program of the LHC, a next-generation accelerator, the Future Circular Collider (FCC), has been proposed. The FCC aims to significantly expand both the energy and luminosity frontiers in order to enhance sensitivity to new physics phenomena. Its operation is planned in two stages: an initial electron--positron collider phase (FCC-ee), followed by a high-energy hadron collider phase (FCC-hh)~\cite{CERN:FCC}.

One of the most significant achievements of the LHC was the discovery of the Higgs boson in 2012 by the ATLAS and CMS collaborations~\cite{CERN:StandardModel}. As the only fundamental scalar particle in the SM, the Higgs boson occupies a unique role, and extensive efforts have since been devoted to measuring its properties with increasing precision. In parallel, numerous theoretical extensions of the SM have been developed to address its shortcomings and unresolved questions. Many of these models predict the existence of exotic Higgs boson decays, in which the Higgs decays into new, previously undiscovered particles~\cite{Curtin:2014cca}. Such particles may appear as long-lived particles (LLPs), characterized by macroscopic lifetimes that lead to distinctive experimental signatures, including displaced vertices (DVs) in particle detectors.

The primary objective of this thesis is to investigate exotic Higgs boson decays into LLPs and to assess the feasibility of studying such scenarios at the FCC-ee. In particular, this work focuses on models in which the LLPs are new long-lived scalar particles. This study represents the first analysis of this type performed within the \textsc{FCCAnalyses} framework, a newly developed computational tool designed to support simulation and analysis efforts for the FCC physics program. To achieve this goal, the thesis is structured into three successive studies. The first step involves defining and characterizing the production and decay of long-lived scalars at the FCC-ee, which establishes the signal signature targeted in the analysis.

\section{The Higgs Boson}

The Higgs boson is the final fundamental particle predicted by the Standard Model (SM) and was discovered in 2012 by the ATLAS and CMS experiments at the Large Hadron Collider (LHC)~\cite{ATLAS:2012yve, CMS:2012qbp}. It is the only elementary particle with spin zero that has been observed to date, making it a unique component of the SM. The Higgs boson carries neither electric charge nor color charge; instead, it interacts with other particles through couplings proportional to their masses.

This particle is associated with the Higgs field, which plays a central role in the generation of mass for all quarks, charged leptons, and the weak gauge bosons. Unlike the other fields in the SM, which have vanishing vacuum expectation values, the Higgs field is characterized by a non-zero vacuum expectation value that permeates all of space. This property gives rise to the Higgs mechanism, through which several fundamental particles acquire mass. At the same time, this mechanism is responsible for the spontaneous breaking of the unified electroweak interaction into the distinct electromagnetic and weak forces observed at low energies.

\subsection{Beyond the Standard Model in Higgs Decays}

In order for a new particle to be produced and detected at a collider, it must interact—directly or indirectly—with particles of the Standard Model (SM). The Higgs boson is particularly well suited for probing such interactions, as its total decay width is extremely small compared to its mass. Current experimental constraints on the Higgs total width therefore allow significant room for couplings to particles beyond the Standard Model (BSM). These interactions can give rise to so-called exotic Higgs decays, in which the Higgs boson decays into new, light BSM particles~\cite{Curtin:2014cca}.

A wide variety of BSM theories predict the existence of exotic Higgs decays, including two-Higgs-doublet models, Little Higgs models, the Next-to-Minimal Supersymmetric Standard Model, and Hidden Valley scenarios. These frameworks address several unresolved issues in particle physics and often introduce new degrees of freedom that couple weakly to the SM~\cite{Curtin:2014cca}. Hidden Valley models, in particular, postulate the existence of an additional sector composed of light states that interact feebly with SM particles through a heavy mediator~\cite{Strassler:2006im}. Such models are motivated, for example, by theories of neutral naturalness, which aim to resolve the hierarchy problem without introducing colored top partners~\cite{Craig:2015pha}. Because this new sector couples only weakly to the SM, it is commonly referred to as a hidden or dark sector. Similar hidden-sector constructions also arise in many models proposed to explain the nature of dark matter~\cite{Krnjaic:2015mbs}.

The mediator connecting the SM and the hidden sector is often referred to as a \emph{portal}. In many scenarios, this portal is realized by a new scalar particle that couples to the Higgs boson. Depending on the model parameters, this scalar can be long-lived, leading to distinctive experimental signatures. A comprehensive overview of exotic Higgs decays, their theoretical motivations, and associated experimental signatures is provided in Ref.~\cite{Curtin:2014cca}, while a more recent review focusing on searches for long-lived particles (LLPs) can be found in Ref.~\cite{Cepeda:2021rql}.

Future collider experiments offer enhanced sensitivity to such scenarios and enable detailed studies of long-lived scalars. Previous studies~\cite{Read:2002hq,AlipourFard:2018lsf} have examined the sensitivity of the proposed FCC-ee collider to LLPs originating from exotic Higgs boson decays. As discussed in Section~5, these analyses indicate a projected peak sensitivity corresponding to an upper limit of
\[
\mathrm{BR}(h \to ss) \sim 5 \times 10^{-5} \quad \text{(95\% CL)}
\]
for scalar masses in the range $2.5$--$50~\mathrm{GeV}$ and decay lengths between $0.01~\mathrm{mm}$ and $10~\mathrm{m}$. These results can be translated into a lower bound on the coupling strength, $k \sim 10^{-3}$, and constraints on the scalar--Higgs mixing angle in the range $\sin\theta \sim 10^{-7}$ to $10^{-4}$ for a scalar mass of $50~\mathrm{GeV}$~\cite{AlipourFard:2018lsf}.

The integrated luminosity at each operational stage of the FCC-ee is determined by the instantaneous luminosity and the duration of data taking. For instance, during the Higgsstrahlung ($Zh$) stage, the expected integrated luminosity is $L = 5~\mathrm{ab}^{-1}$. The corresponding projected event yields are approximately $5 \times 10^{12}$ Z bosons at $\sqrt{s}=91~\mathrm{GeV}$, $10^{8}$ W-boson pairs at $160~\mathrm{GeV}$, $10^{6}$ Higgs bosons at $240~\mathrm{GeV}$, and $10^{6}$ $\tau^+\tau^-$ pairs at $350~\mathrm{GeV}$. These exceptionally large data samples will enable measurements of unprecedented precision, significantly surpassing those achievable at current colliders. Any observed deviations from SM predictions would provide indirect evidence for new physics at energy scales beyond the direct reach of existing experiments~\cite{Abada:2019zxq}.

While the primary physics goals of the FCC-ee include high-precision measurements of the Higgs boson, electroweak gauge bosons, and the top quark, the clean experimental environment and high statistics of electron--positron collisions also offer powerful opportunities for the direct discovery of BSM phenomena. In particular, theories predicting new particles with very small couplings to the SM can be efficiently probed. Such particles may appear experimentally as LLPs. Notable LLP scenarios with strong discovery potential at the FCC-ee include Heavy Neutral Leptons (HNLs), Axion-Like Particles (ALPs), and a wide class of exotic Higgs boson decays to long-lived states

\subsection{Dark Matter}

The scotogenic model extends the Standard Model (SM) by introducing additional fermionic and scalar fields, namely the fermions $N_{1,2,3}$ and the scalars $S$, $P$, and $H^\pm$. All of these new particles are odd under an imposed $Z_2$ symmetry, which guarantees the stability of the lightest $Z_2$-odd particle and allows it to serve as a viable dark matter (DM) candidate.

In this work, we consider the commonly studied scenario in which the lightest fermion, $N_1$, constitutes the DM particle, while $N_2$ is assumed to be nearly mass-degenerate with $N_1$. This mass configuration is phenomenologically favorable, as it enables consistency with both the observed DM relic abundance and the experimental bounds on the lepton-flavor-violating decay $\mu \to e \gamma$.

The DM relic abundance is conventionally expressed in terms of the present-day DM density relative to the critical density, $\Omega$, multiplied by the squared reduced Hubble parameter, $\hat h^2$. Its theoretical prediction is given by~\cite{Kubo:2006yx,Jungman:1995df}
\begin{eqnarray} \label{omega}
	\Omega \hat h^2 &\,=\,& \frac{1.07\times10^9\;x_f\,{\rm GeV}^{-1}}{\sqrt{g_*}\;m_{\rm Pl}\,\bigl[a_{\rm eff}+3(b_{\rm eff}-a_{\rm eff}/4)/x_f\bigr]} ~,\nonumber \\ 
	x_f &\,=\,& \ln\frac{0.191\,\bigl(a_{\rm eff}+6 b_{\rm eff}/x_f\bigr)M_1\,m_{\rm Pl}}{\sqrt{g_*\,x_f}} ~,
\end{eqnarray}
where $g_*$ denotes the effective number of relativistic degrees of freedom at temperatures below the freeze-out temperature $T_f = M_1/x_f$, and $m_{\rm Pl} = 1.22 \times 10^{19}~\mathrm{GeV}$ is the Planck mass.

The coefficients $a_{\rm eff}$ and $b_{\rm eff}$ arise from the non-relativistic expansion of the effective co-annihilation cross section,
$\sigma_{\rm eff} v_{\rm rel} = a_{\rm eff} + b_{\rm eff} v_{\rm rel}^2$,
in terms of the relative velocity $v_{\rm rel}$ of the annihilating particles in their center-of-mass frame. The effective cross section is defined as
\[
\sigma_{\rm eff} = \frac{1}{4}\left(\sigma_{11} + \sigma_{12} + \sigma_{21} + \sigma_{22}\right),
\]
where the individual cross sections $\sigma_{ij}$ ($i,j=1,2$) are given by
\[
\sigma_{ij} = \sigma_{N_i N_j \to \ell_i^- \ell_j^+} + \sigma_{N_i N_j \to \nu_i \nu_j}.
\]

These annihilation cross sections have been calculated in Refs.~\cite{Ho:2013hia,Ho:2013spa} and originate from tree-level $t$- and $u$-channel processes mediated by the charged scalar $H^\pm$ or the neutral scalars $(\mathcal{S}, P)$, depending on whether the final states consist of charged leptons or neutrinos.

In order to constrain the model parameter space using the observed DM relic abundance, our analysis assumes that the DM particle and the scalar states are not mass-degenerate. This assumption suppresses scalar co-annihilation contributions and isolates the fermionic annihilation channels relevant for determining the relic density.

\subsection{Hidden Abelian Higgs Model}

One of the simplest extensions of the Standard Model (SM) consists of introducing an additional Abelian $U(1)$ gauge symmetry associated with a hidden sector. This symmetry is spontaneously broken by the vacuum expectation value of a dark Higgs field. As a result, a new massive gauge boson, commonly denoted $Z_D$, emerges and can kinetically mix with the SM $U(1)_Y$ gauge boson. In parallel, the dark Higgs scalar $S$ (or $h_S$) can mix with the SM Higgs doublet through the Higgs portal interaction $|H|^2|S|^2$.

Depending on the dominant mixing mechanism, this framework allows for a variety of exotic Higgs boson decay channels. If kinetic mixing dominates, decays such as $h \to Z Z_D$ may occur, while Higgs mixing can lead to decays like $h \to Z_D Z_D$ or $h \to S S$. The scalar $S$ inherits its decay modes from the SM Higgs boson, whereas the dark gauge boson $Z_D$ couples proportionally to gauge charge, often resulting in lepton-rich final states. More intricate cascade decays are also possible, for example
\[
h \to S S \to Z_D Z_D Z_D Z_D,
\]
which can give rise to striking signatures with up to eight charged leptons in the final state.

The phenomenology of this model has been extensively explored in the literatures. 
 In addition, the Exotic Higgs Decays Working Group presented a comprehensive survey of possible exotic Higgs decay modes, including scenarios involving a dark vector and a dark Higgs field~\cite{Curtin:2014cca}.

To facilitate experimental investigations of these channels, a dedicated \textsc{MadGraph5} model has been constructed to implement this SM extension. The model is developed using \texttt{FeynRules}~2.0.23 and is based on the Hidden Abelian Higgs Model (HAHM), but includes several important improvements to ensure internal consistency and numerical accuracy. These modifications include the correction of typographical errors, a fully self-consistent determination of all mixing angles from the physical mass eigenvalues $M_Z$, $M_{Z_D}$, $M_h$, and $M_{h_S}$, and the elimination of the small-mixing approximation.

The mixing angles are defined such that, for small Higgs or kinetic mixing ($\ll 1$), the corresponding angles remain small regardless of the relative mass hierarchy between the SM Higgs or $Z$ boson and the dark scalar $h_S$ or dark gauge boson $Z_D$. Furthermore, SM relations modified by $Z$--$Z_D$ mixing are treated consistently using two alternative schemes: either by shifting the $W$-boson mass or by redefining the Weinberg angle.

The model is additionally combined with the Higgs Effective Theory framework, allowing for the inclusion of effective $hgg$ and $h\gamma\gamma$ operators. By appropriately choosing the mixing parameters, the model can also reproduce NMSSM-like decay topologies such as $h \to S S \to 4b$ by suppressing kinetic mixing, or pure kinetic-mixing phenomenology by setting the Higgs mixing to zero (while avoiding exactly vanishing values to maintain numerical stability).

In the scotogenic model considered in this study, the production of charged scalars at an electron--positron collider proceeds via the process $e^+(p_+)e^-(p_-)\to H^+H^-$. The corresponding amplitude receives contributions from tree-level diagrams involving photon ($\gamma$), $Z$-boson, and fermion $N_{1,2,3}$ exchange. Neglecting the electron mass, the total cross section is given by~\cite{Ho:2013spa}
\begin{eqnarray} \label{sigma_ee2HH}
	&&\sigma_{e^+e^-\to H^+ H^-} =
	\frac{\pi\alpha^2\beta^3}{3 s} +
	\frac{\alpha}{12} \frac{\bigl(g_L^2+g_L g_R\bigr)\beta^3}{s-m_Z^2} +
	\frac{\bigl(g_L^4+g_L^2g_R^2\bigr)\beta^3 s}{96\pi\bigl(s-m_Z^2\bigr)^2}
	\nonumber \\ &&+
	\sum_k \frac{|{\cal Y}_{1k}|^4}{64\pi\,s}
	\Biggl(w_k\,\ln\frac{w_k+\beta}{w_k-\beta}-2\beta\Biggr)
	+ \Biggl[ \frac{\alpha}{16 s} + \frac{g_L^2}{64\pi\,\bigl(s-m_Z^2\bigr)}\Biggr]\nonumber\\&&
	\sum_k |{\cal Y}_{1k}|^2
	\biggl[\bigl(w_k^2-\beta^2\bigr)\ln\frac{w_k+\beta}{w_k-\beta}-2\beta\,w_k\biggr]
	\nonumber \\ &&+
	\sum_{j,\,k>j}
	\frac{|{\cal Y}_{1j}{\cal Y}_{1k}|^2}{64\pi\,s}
	\Biggl(
	\frac{w_j^2-\beta^2}{w_j-w_k}\ln\frac{w_j+\beta}{w_j-\beta}
	+ \frac{w_k^2-\beta^2}{w_k-w_j}\ln\frac{w_k+\beta}{w_k-\beta}
	-2\beta
	\Biggr),
\end{eqnarray}
where $j,k=1,2,3$, $s=(p_+ + p_-)^2$, $\alpha = e^2/(4\pi)$,
$\beta = \sqrt{1-4m_H^2/s}$, and
\[
w_k = 1 + \frac{2M_k^2}{s} - \frac{2m_H^2}{s}\beta.
\]
For the numerical analysis, we use the effective input values $\alpha = 1/128$, $g = 0.6517$, and $s_w^2 = 0.23146$~\cite{pdg}.

\subsection{Constraints}

Dedicated searches for exotic Higgs boson decays into long-lived scalar particles, followed by various Standard Model (SM) decay modes, are currently being pursued at several collider experiments. To date, no evidence for such scalars has been observed, and upper limits have been placed on the branching ratio $\mathrm{BR}(h \to ss)$. Figure~4 summarizes the existing exclusion limits for mean proper lifetimes ranging from $0.1~\mathrm{mm}$ to $10~\mathrm{m}$ and scalar masses in the interval $m_s = 30$--$40~\mathrm{GeV}$. The curves shown correspond to different analyses performed by the ATLAS, CMS, and LHCb collaborations, each probing complementary final states and long-lived particle (LLP) signatures.

These searches target a wide variety of experimental signatures, including tracks with large impact parameters and displaced vertices (DVs) reconstructed in the inner tracking detector (ID), displaced hadronic jets or photons in the calorimeters, as well as displaced jet or lepton signals in the muon system (MS). A more detailed discussion of LLP signatures is provided in Section~6. All limits are reported at the 95\% confidence level (CL), which is the standard criterion for exclusion in particle physics analyses~\cite{Cepeda:2021rql,ROOT:ReferenceV626}.

Future collider facilities offer the opportunity to extend the sensitivity to long-lived scalars well beyond current limits. In particular, a previous study~\cite{AlipourFard:2018lsf} evaluated the projected reach of the proposed FCC-ee collider for LLPs originating from exotic Higgs boson decays. As described in Section~5, the FCC-ee is expected to achieve a peak sensitivity corresponding to an upper limit of
\[
\mathrm{BR}(h \to ss) \sim 5 \times 10^{-5} \quad \text{(95\% CL)},
\]
for scalar masses in the range $2.5$--$50~\mathrm{GeV}$ and decay lengths between $0.01~\mathrm{mm}$ and $10~\mathrm{m}$. These bounds can be translated into a lower limit on the effective coupling, $k \sim 10^{-3}$, and constraints on the scalar--Higgs mixing angle of $\sin\theta \sim 10^{-7}$ to $10^{-4}$ for a scalar mass of $50~\mathrm{GeV}$~\cite{AlipourFard:2018lsf}.

Throughout this work, we adopt the normal ordering (NO) for neutrino masses. The leptonic mixing angles, Dirac CP-violating phase, and mass-squared differences $|\Delta m_{31}^2|$ and $\Delta m_{21}^2$ are fixed using the global neutrino oscillation fit reported in Ref.~\cite{deSalas:2020pgw}. This analysis imposes the constraint
\[
32.0 < R_m \equiv \frac{|\Delta m^2_{31}|}{\Delta m^2_{21}} < 36.0
\]
on the parameter space at the 90\% CL. In addition, cosmological observations from the \textit{Planck} satellite, baryon acoustic oscillations (BAO), measurements of $H(z)$, and Type Ia supernovae yield a stringent $2\sigma$ upper bound on the sum of neutrino masses,
\[
\sum_i m_i < 0.12~\mathrm{eV}.
\]
Constraints from neutrinoless double beta decay experiments further restrict the effective Majorana mass to
\[
|\langle m\rangle_{ee}| < 0.06\text{--}0.2~\mathrm{eV} \quad (95\%~\mathrm{CL}),
\]
where $\langle m\rangle_{ee} = m_1 U_{e1}^2 + m_2 U_{e2}^2 + m_3 U_{e3}^2$~\cite{EXO-200:2019rkq,KamLAND-Zen:2016pfg,GERDA:2020xhi}.

The Yukawa interactions involving the charged scalar $H^\pm$, as introduced in Eq.~(\ref{LN}), induce lepton-flavor-violating (LFV) processes at the one-loop level. A detailed discussion of these contributions is provided in Ref.~\cite{Toma:2013zsa}. Current experimental upper limits on LFV branching ratios are given by $\mathrm{BR}(\mu \to e\gamma) < 4.2 \times 10^{-13}$~\cite{MEG:2016leq}, $\mathrm{BR}(\tau \to e\gamma) < 3.3 \times 10^{-8}$, and $\mathrm{BR}(\tau \to \mu\gamma) < 4.4 \times 10^{-8}$~\cite{BaBar:2009hkt}, with the most stringent constraint arising from $\mu \to e\gamma$. The corresponding expressions within the scotogenic framework can be found in Ref.~\cite{Toma:2013zsa}. The flavor-conserving counterparts of these processes contribute to the anomalous magnetic moments of charged leptons, leading to~\cite{Ma:2001mr}
\begin{eqnarray} \label{g-2}
	\Delta a_{\ell_i} = 
	\frac{-m_{\ell_i}^2}{16\pi^2 m_H^2}
	\sum_k |Y_{ik}|^2\,
	\mathcal{F}\!\left(\frac{M_k^2}{m_H^2}\right).
\end{eqnarray}

Among these observables, the muon anomalous magnetic moment provides the strongest constraint. The current discrepancy between the experimental measurement and the SM prediction is
\[
a_\mu^{\mathrm{exp}} - a_\mu^{\mathrm{SM}} = (2.51 \pm 0.59) \times 10^{-9},
\]
as reported in Ref.~\cite{Aoyama:2020ynm}.

Direct detection constraints on dark matter (DM), arising from elastic scattering of $N_1$ off nucleons via one-loop Higgs exchange, have been analyzed in Ref.~\cite{Ibarra:2016dlb}. To avoid the stringent bounds from direct detection experiments such as XENON1T and PandaX-4T~\cite{XENON:2018voc,PandaX-4T:2021bab}, we follow Refs.~\cite{Ibarra:2016dlb,Liu:2022byu} and set $\lambda_{3,4} = 0.01$. This choice leads to
\[
m_0 \simeq m_{H^\pm} + \tfrac{1}{2}\lambda_4 v^2 \simeq m_{H^\pm} + 350~\mathrm{GeV}.
\]
While direct detection constraints were not included in earlier analyses~\cite{Ho:2013spa}, they are incorporated in this work together with the latest global neutrino oscillation data~\cite{deSalas:2020pgw}.

In the following section, we present the numerical results of our analysis. The leptonic mixing angles $\theta_{12}$, $\theta_{23}$, $\theta_{13}$ and the Dirac phase $\delta$ are fixed to their best-fit values from Ref.~\cite{deSalas:2020pgw}. A comprehensive scan is then performed over the remaining model parameters, including the scalar masses $m_H$ and $m_0$, the singlet fermion masses $M_{1,2,3}$, and the Yukawa couplings $Y_{1,2,3}$ defined in Eq.~(\ref{123}). For light DM masses $M_1 < 100~\mathrm{GeV}$, the parameter space is predominantly excluded by LFV constraints~\cite{Vicente:2014wga} and LHC searches~\cite{ATLAS:2019lng,CMS:2020bfa}. Additional bounds on scalar masses arise from measurements of the $W$ and $Z$ boson widths and the absence of new particle signals at $e^+e^-$ colliders, leading to the following constraints~\cite{Arhrib:2012ia,Cao:2007rm,Pierce:2007ut}:
\begin{eqnarray}
	m_{H^\pm} + m_{\mathcal{S},\mathcal{P}} &>& m_{W^\pm}, \nonumber\\
	m_{H^\pm} &>& 70~\mathrm{GeV}, \nonumber\\
	m_{\mathcal{S}} + m_{\mathcal{P}} &>& m_Z.
\end{eqnarray}

\section{Simulation Setting}

Simulations of particle collisions and detector responses play a central role in modern particle physics, serving both precision measurements and searches for physics beyond the Standard Model (BSM). Simulated data, grounded in theoretical predictions, provide the basis for testing hypotheses against experimental observations. In feasibility and sensitivity studies, simulations are indispensable for accurately modeling particle kinematics and detector effects, enabling reliable estimates of signal sensitivity and offering guidance for detector design and optimization.

This analysis is performed within the \textsc{FCCAnalyses} framework~\cite{FCCAnalyses:GitHub}. Event generation and detector simulation are carried out using the software packages \textsc{MadGraph}~\cite{Alwall:2014hca}, \textsc{PYTHIA}~\cite{Bierlich:2022Pythia83}, and \textsc{DELPHES}~\cite{Ovyn:2009txa}, with event outputs stored in the \textsc{EDM4HEP} format~\cite{EDM4hep:GitHub}. The subsequent analysis is implemented in Python and primarily relies on the \texttt{RDataFrame} interface of the \textsc{ROOT} data analysis framework~\cite{ROOT:ReferenceV626}. Within \textsc{FCCAnalyses}, the analysis workflow is organized into three sequential stages: the first processes the simulated samples and extracts the relevant physics variables; the second defines the event selection and specifies the quantities to be histogrammed; and the final stage produces the corresponding distributions. The remainder of this section provides a detailed description of the simulation tools used in each step.

An $e^-e^+$ collision event is inherently complex, involving multiple stages of particle interactions. For simulation purposes, the event generation is therefore decomposed into several successive components. These include the hard scattering process of the initial electron--positron collision, followed by parton showering and hadronization to model the formation of final-state particles, and finally the detector simulation to reproduce the experimental response. Each of these stages is computed using Monte Carlo (MC) generators, which employ stochastic sampling techniques to obtain numerical solutions for the underlying physical processes.

At the level of the hard scattering, the cross section and kinematic properties of the signal process are calculated for the transition from the initial $e^-e^+$ state to an $N$-particle final state. At this stage, colored particles such as quarks, antiquarks, and gluons are treated as free, non-interacting objects. In this study, these calculations are performed using \textsc{MadGraph}, a widely used MC generator capable of simulating both SM and BSM processes. \textsc{MadGraph} computes matrix elements at leading order, and can also incorporate next-to-leading-order corrections to improve the theoretical precision~\cite{Alwall:2014hca}.

Before and after the hard scattering, the incoming and outgoing charged particles may emit additional radiation, giving rise to initial- and final-state radiation. These effects are modeled using parton shower algorithms. Since free colored particles are not observed experimentally, all quarks and gluons produced either in the hard process or during the parton shower must undergo hadronization to form color-neutral hadrons. This step is handled using \textsc{PYTHIA}, which supplements the hard-scattering events generated by \textsc{MadGraph} with parton showering, hadronization, and other strong-interaction phenomena~\cite{Bierlich:2022Pythia83}.

The final stage of the simulation chain models the interaction of the produced particles with the detector material. This analysis employs the \textsc{DELPHES} framework with an IDEA detector configuration to perform a fast, parametric detector simulation. In such simulations, the detector response is described using parameterized resolutions and efficiencies that depend on the particle type and detector region. This approach contrasts with full detector simulations, in which particles are propagated step-by-step through the detector geometry and material, and physical interactions are modeled in detail using, for example, the Bethe--Bloch formalism. The \textsc{DELPHES} simulation includes a tracking system in a magnetic field, electromagnetic and hadronic calorimeters, and a muon system, and provides reconstructed observables such as leptons, jets, and missing transverse energy~\cite{Ovyn:2009txa}.

\section{Results and Discussion}
\subsection{Production and Decay at FCC-ee}

The signal process studied in this thesis is
$h \rightarrow ss \rightarrow b\bar{b}b\bar{b}$, corresponding to the decay of the Higgs boson into a pair of scalar particles, each of which subsequently decays into a bottom--antibottom quark pair. The choice of a $b$-quark final state is motivated by the fact that it represents the dominant decay mode for scalar masses above $10~\mathrm{GeV}$, as illustrated in Figure~3.

\begin{figure}[htbp]
	\centering
	\includegraphics[width=0.75\textwidth]{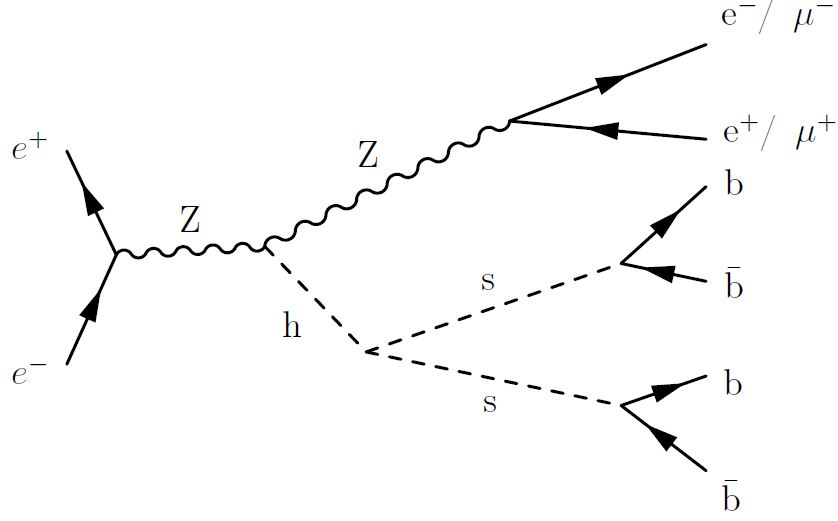}
	\caption{Feynman diagram of the complete signal process
		$h \rightarrow ss \rightarrow b\bar{b}b\bar{b}$
		produced at the FCC-ee via $e^{+}e^{-} \rightarrow Zh$, with the $Z$ boson decaying into $e^{+}e^{-}$ or $\mu^{+}\mu^{-}$.}
	\label{fig:Fynman_Diagram}
\end{figure}

At the FCC-ee, Higgs bosons are primarily produced in association with a $Z$ boson at a center-of-mass energy of $\sqrt{s} = 240~\mathrm{GeV}$. In the simulated events, the $Z$ boson is required to decay leptonically into either an electron or a muon pair, according to
$Z \rightarrow e^{+}e^{-}$ or $Z \rightarrow \mu^{+}\mu^{-}$. The full signal production process at the FCC-ee can therefore be written as
\\
\begin{equation}
	\label{eq81}
	e^{+}e^{-} \rightarrow Zh,\quad
	Z \rightarrow e^{+}e^{-}\ \text{or}\ \mu^{+}\mu^{-},\quad
	h \rightarrow ss \rightarrow b\bar{b}b\bar{b}.
\end{equation}
\\
This process is illustrated by the Feynman diagram shown in Figure~\ref{fig:Fynman_Diagram}. In the diagram, the incoming electron and positron are represented by the leftmost lines. The production of the $Z$ boson and the Higgs boson proceeds via an intermediate $Z$ boson, depicted in the central part of the diagram. The upper branches correspond to the leptonic decay of the $Z$ boson into an $e^{+}e^{-}$ or $\mu^{+}\mu^{-}$ pair, while the rightmost branches show the decay of each scalar particle $s$ into a $b\bar{b}$ quark pair.
\\
\\
\subsection{Simulating the Signal Process}

In this analysis, the normal ordering (NO) of neutrino masses is assumed. The leptonic mixing angles, the Dirac CP-violating phase, and the mass-squared differences $|\Delta m^2_{31}|$ and $\Delta m^2_{21}$ are constrained by a variety of experimental measurements. For the numerical evaluation, we adopt the results of the global neutrino oscillation data fit presented in Ref.~\cite{deSalas:2020pgw}. This fit imposes the condition
\[
32.0 < R_m \equiv \frac{|\Delta m^2_{31}|}{\Delta m^2_{21}} < 36.0
\]
at the 90\% confidence level. In addition, cosmological observations from the \textit{Planck} satellite, baryon acoustic oscillation (BAO) data, measurements of $H(z)$, and Type Ia supernovae constrain the sum of neutrino masses to $\sum m_i < 0.12~\mathrm{eV}$ at the $2\sigma$ level. Constraints from neutrinoless double beta decay experiments further limit the effective Majorana mass to
$|\langle m\rangle_{ee}| = 0.06$--$0.2~\mathrm{eV}$ at 95\% CL~\cite{EXO-200:2019rkq,KamLAND-Zen:2016pfg,GERDA:2020xhi}, where
$\langle m\rangle_{ee}= m_1 U_{e1}^2 + m_2 U_{e2}^2 + m_3 U_{e3}^2$.

To generate the signal process defined in \ref{eq81}, the \textsc{HAHM-MG5 Model-v3}~\cite{Curtin:2014cca} is employed within \textsc{MadGraph}. This model corresponds to the Hidden Abelian Higgs Model (HAHM), which introduces an additional scalar particle as well as a vector boson commonly referred to as the dark photon. The model is characterized by four input parameters: the dark photon mass $m_{Z_D}$, the kinetic mixing parameter $\epsilon$, the scalar mass $m_s$, and the scalar--Higgs coupling constant.

In order to isolate the SM$+$scalar scenario, the dark photon is effectively decoupled by assigning it a large mass, $m_{Z_D}=1000~\mathrm{GeV}$, and a very small kinetic mixing, $\epsilon = 10^{-10}$. To produce long-lived scalar particles, the total decay width of the scalar is explicitly specified in the \textsc{MadGraph} process card. For scalars decaying into a $b\bar{b}$ final state, the total decay width is given by~\cite{Curtin:2014cca}
\\
\begin{equation}
	\label{eq82}
	\Gamma_s
	=
	\sin^2\theta \,
	\frac{3}{0.9 \times 8\pi} \,
	\frac{m_s m_b^2}{v_h^2}
	\left(1 - \frac{4 m_b^2}{m_s^2}\right)^{3/2}.
\end{equation}
\\
Here, the branching ratio $\mathrm{BR}(s \rightarrow b\bar{b})$ is fixed to 0.9, as shown in the left panel of Figure~3, and the expression for the partial decay width at leading order 
The numerical values used are $m_b = 4.2~\mathrm{GeV}$ for the $b$-quark mass, $v_h = 246~\mathrm{GeV}$ for the Higgs vacuum expectation value, and the number of colors $N_c = 3$. This procedure allows precise control over the scalar lifetime.

All signal samples are generated with 10{,}000 events each. The scalar masses, mixing angles, and corresponding decay widths used in the analysis are summarized in Table~\ref{tab:scalar_widths}. Two representative scalar masses, $m_s = 20~\mathrm{GeV}$ and $m_s = 60~\mathrm{GeV}$, are considered, covering distinct regions of the parameter space where decays into $b\bar{b}$ pairs are kinematically allowed. For each mass point, the mixing angles are chosen as $\sin\theta = 10^{-5}$, $10^{-6}$, and $10^{-7}$, corresponding to proper decay lengths ranging approximately from $1~\mathrm{mm}$ to $10~\mathrm{m}$. The scalar--Higgs coupling constant is fixed to $\kappa = 0.001$ for all signal samples, yielding a branching ratio of order
$\mathrm{BR}(h \rightarrow ss) \sim 10^{-4}$. This value is below current experimental limits from LHC searches and lies well within the projected sensitivity of the FCC-ee~\cite{AlipourFard:2018lsf}.
\\
\begin{table}[htbp]
	\centering
	\caption{Scalar masses, mixing angles, and corresponding decay widths used in the analysis.}
	\label{tab:scalar_widths}
	\begin{tabular}{c c c}
		\hline\hline
		Mass of Scalar & Mixing angle & Width of Scalar \\
		$m_s~[\mathrm{GeV}]$ & $\sin\theta$ & $\Gamma_s~[\mathrm{GeV}]$ \\
		\hline
		20 & $1 \times 10^{-5}$ & $5.779 \times 10^{-14}$ \\
		20 & $1 \times 10^{-6}$ & $5.779 \times 10^{-16}$ \\
		20 & $1 \times 10^{-7}$ & $5.779 \times 10^{-18}$ \\
		\hline\hline
	\end{tabular}
\end{table}
\\
The complete simulation chain is executed sequentially using \textsc{MadGraph}~\cite{Alwall:2014hca} v3.4.1, \textsc{PYTHIA}~\cite{Bierlich:2022Pythia83} v8.303, and \textsc{DELPHES}~\cite{Ovyn:2009txa} v3.4.2. For each signal sample, the process and parameter definitions listed in Table~\ref{tab:scalar_widths} are specified in a \textsc{MadGraph} process card, which serves as input to generate the hard-scattering events. The resulting events are subsequently passed to \textsc{PYTHIA} for parton showering and hadronization. Finally, detector effects are simulated using \textsc{DELPHES} with the \texttt{spring2021} IDEA detector card, which defines the detector geometry and response for the fast parametric simulation.

\subsection{Validation of Generated Quantities}

The validation of the signal samples is performed by examining selected truth-level quantities, that is, kinematic variables evaluated prior to the inclusion of detector effects. Particular attention is given to the decay lengths and time distributions of the long-lived scalar particles. These observables provide an essential cross-check of the correct implementation of the model parameters and the expected lifetime behavior of the scalars.

The results are presented in the form of histograms, where each bin is filled separately for each scalar particle. The histograms include overflow entries, such that events with values exceeding the upper bound of the x-axis are accumulated in the final bin. This treatment accounts for the enhanced population observed in the last bin of some distributions.

Most of the histograms presented in the following sections share a common labeling scheme. In the upper-left corner, the first line indicates the center-of-mass energy, $\sqrt{s} = 240~\mathrm{GeV}$, while the second line specifies the integrated luminosity, $L = 5~\mathrm{ab}^{-1}$. The third line displays the signal process defined in\ref{eq81}, and the final line indicates that no event selection has been applied.

In the legend, the individual signal samples are identified by their scalar mass $m_s$ and mixing angle $\sin\theta$. The x-axis corresponds to the observable under study, whereas the y-axis shows the number of events per bin width, displayed on a logarithmic scale. Event yields are normalized to the corresponding production cross sections, listed in Table~8, and scaled to the assumed integrated luminosity.

Finally, each plot title indicates that the simulations are performed for the FCC-ee using the \textsc{Delphes} detector simulation within the \textsc{FCCAnalyses} framework.

\subsubsection{Generated Mass and Number of Scalars}

\begin{figure}[htbp]
	\centering
	\includegraphics[width=0.75\textwidth]{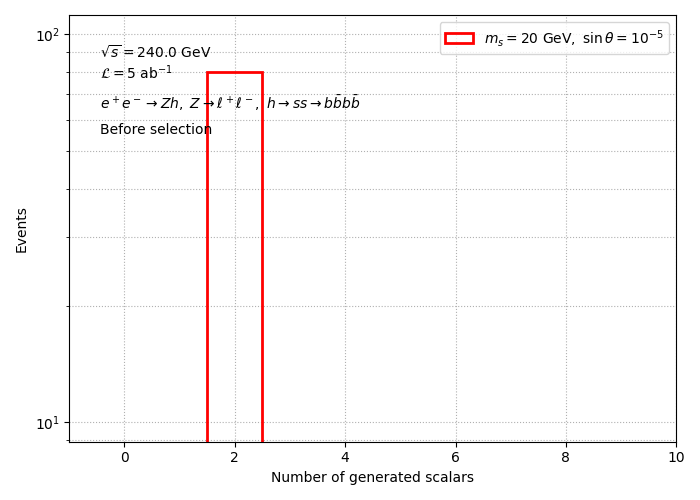}
	\caption{The number of generated scalars in each event for the signal sample
		$m_s = 20~\mathrm{GeV}$ and $\sin\theta = 1\times10^{-5}$.}
	\label{fig:no_generated_scalar}
\end{figure}

\begin{figure}[htbp]
	\centering
	\includegraphics[width=0.75\textwidth]{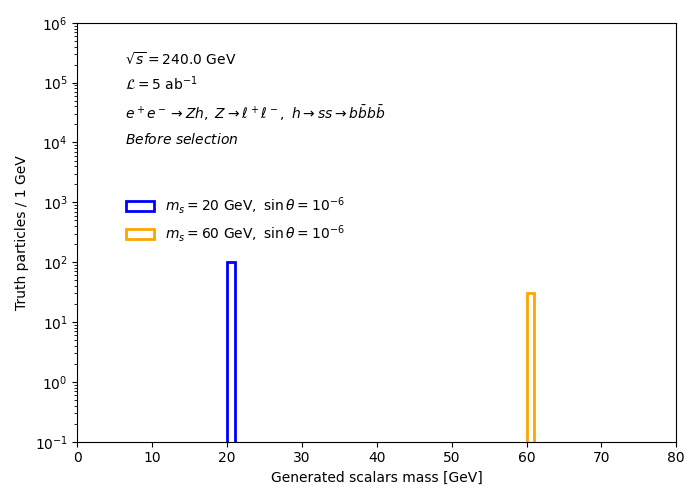}
	\caption{Generated mass of the scalars for signal samples with $\sin\theta = 1\times10^{-6}$ and
		masses $m_s = 20~\mathrm{GeV}$ and $m_s = 60~\mathrm{GeV}$.}
	\label{fig:generated_scalar_mass}
\end{figure}

As an initial validation step, the number of scalar particles produced per event is examined. Figure~\ref{fig:no_generated_scalar} displays the distribution of the number of generated scalars for the signal sample with $m_s = 20~\mathrm{GeV}$ and $\sin\theta = 10^{-5}$. According to the signal definition given in Eq.~(\ref{eq81}), each event contains exactly two scalar particles. This expectation is confirmed by Figure~\ref{fig:no_generated_scalar}, which shows that all events populate the bin corresponding to two scalars, demonstrating that the event generation has been implemented correctly.
\\
\\
The next quantity validated is the generated mass of the scalar particles. As summarized in Table~1, the signal samples are produced with two distinct scalar masses, $m_s = 20~\mathrm{GeV}$ and $m_s = 60~\mathrm{GeV}$. Figure~\ref{fig:generated_scalar_mass} shows the generated scalar mass distributions for samples with a common mixing angle of $\sin\theta = 10^{-6}$ and scalar masses of $m_s = 20~\mathrm{GeV}$ (blue) and $m_s = 60~\mathrm{GeV}$ (orange). The distributions exhibit sharp peaks at the corresponding input mass values, with no entries in the remaining bins, confirming that the scalar masses are generated as expected.
\\
\\
It should be noted that both scalar particles produced in each event are included in the distribution shown in Figure~\ref{fig:generated_scalar_mass}. Consequently, the histogram is filled twice per event, once for each scalar. This explains why the maximum bin content for the $m_s = 20~\mathrm{GeV}$, $\sin\theta = 10^{-6}$ sample (blue) is twice that observed for the corresponding $m_s = 20~\mathrm{GeV}$, $\sin\theta = 10^{-5}$ sample (red) shown previously.

\subsubsection{Generated Decay Lengths}

To validate the implementation of long-lived scalar particles in the simulation, the decay lengths $L_{xyz}$ of the scalars are examined. This quantity corresponds to the distance traveled by a scalar particle in the laboratory frame before decaying into a $b\bar{b}$ pair. The decay length is defined as
\\
\begin{equation}
	\label{eqpage33}
	L_{xyz}^{\mathrm{truth}} = \sqrt{\Delta x^{2} + \Delta y^{2} + \Delta z^{2}} \, ,
\end{equation}
\\
where $\Delta x$, $\Delta y$, and $\Delta z$ denote the spatial separations between the production vertex of the scalar $s$ and the decay vertex of the resulting $b\bar{b}$ pair in the $x$, $y$, and $z$ directions, respectively.

\begin{figure}[htbp]
	\centering
	\includegraphics[width=0.75\textwidth]{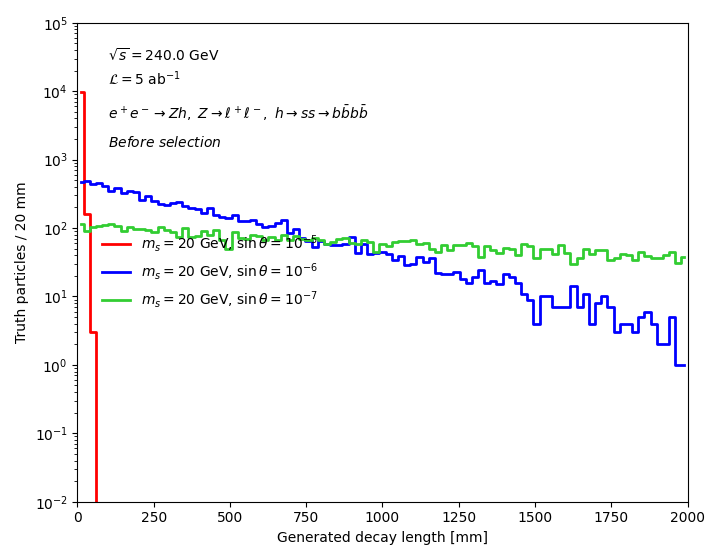}
	\caption{Generated decay lengths, $L_{xyz}^{\mathrm{gen}}$, for all signal samples.}
	\label{fig:decay_length_20}
\end{figure}

The decay length distributions for all signal samples are shown in Figure~\ref{fig:decay_length_20}. The horizontal axis spans the interval from $0$ to $2000~\mathrm{mm}$, corresponding to the region from the $e^{+}e^{-}$ interaction point to the outer radius of the IDEA drift chamber, $R_{\text{out}} = 2~\mathrm{m}$, as illustrated in Figure~9. For signal samples with a mixing angle of $\sin\theta = 10^{-7}$, both for $m_s = 20~\mathrm{GeV}$ (green) and $m_s = 60~\mathrm{GeV}$ (light purple), more than 99\% of the decays fall into the overflow region. This indicates that, for these parameter choices, the scalars are sufficiently long-lived that a large fraction of decays occur outside the tracking volume of the IDEA detector.

For the mixing angle $\sin\theta = 10^{-6}$, the signal sample with the larger scalar mass, $m_s = 60~\mathrm{GeV}$ (orange), exhibits a shorter decay length compared to the sample with $m_s = 20~\mathrm{GeV}$ (blue). This behavior is consistent with the expectations discussed in Section~2.4.2, where the decay width of a particle depends on both the matrix element and the available phase space. For identical coupling strengths, a larger scalar mass leads to an increased decay width and consequently a reduced decay length.

Finally, the dependence of the decay length on the mixing angle is evident when comparing samples with the same scalar mass. Within each mass group, the decay length increases as the mixing angle decreases, in agreement with the scaling behavior described in Eq.~(8.2).
\\
\\
\subsubsection{Generated Time Distributions}

The lifetime properties of the scalar particles are further examined in their rest frame. The measured decay length $L_{xyz}$ is related to the proper lifetime through
\begin{equation}
	L_{xyz} = c\,\tau\,\gamma ,
\end{equation}
where $\tau$ denotes the proper lifetime and $\gamma = E_s/m_s$ is the Lorentz boost factor of the scalar. The proper lifetime can therefore be expressed as
\begin{equation}
	\tau = \frac{L_{xyz}}{c\,\gamma}.
\end{equation}

\begin{figure}[htbp]
	\centering
	\includegraphics[width=0.75\textwidth]{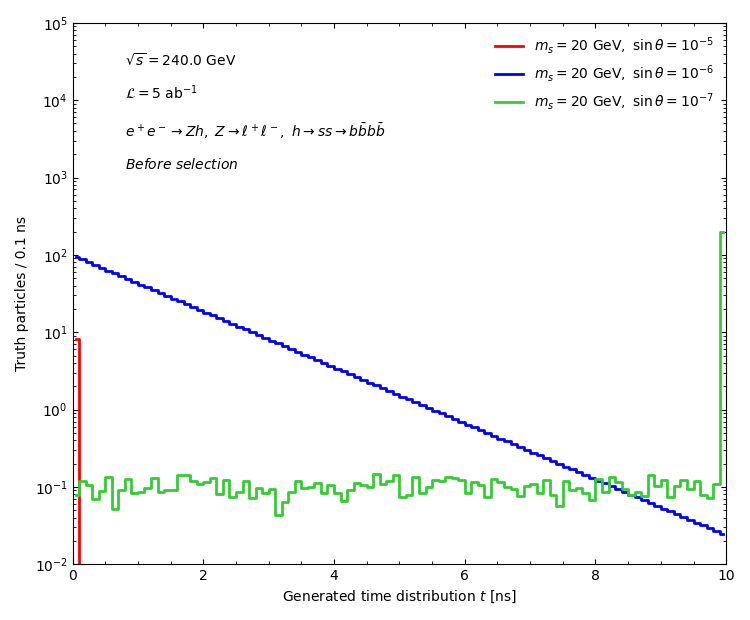}
	\caption{Distribution of the lifetimes for all signal samples.}
	\label{fig:ditribution_of_lifetimes}
\end{figure}

Figure~\ref{fig:ditribution_of_lifetimes} presents the lifetime distributions of the scalar particles for the samples with $m_s = 20~\mathrm{GeV}$ (Fig.~15a) and $m_s = 60~\mathrm{GeV}$ (Fig.~15b). As expected, smaller mixing angles correspond to longer lifetimes. In agreement with the decay-length distributions discussed previously, a fixed mixing angle results in a shorter lifetime for the heavier scalar mass. This effect is particularly evident for $\sin\theta = 10^{-6}$, where the lifetime distribution for $m_s = 20~\mathrm{GeV}$ (blue) extends up to approximately $8~\mathrm{ns}$, while for $m_s = 60~\mathrm{GeV}$ (orange) the distribution falls off around $1~\mathrm{ns}$.
\\
\\
As an additional validation step, the mean proper lifetime obtained from the simulated distributions is compared with the theoretically calculated value. The lifetime spectra are fitted with an exponential function, following Eq.~(2.5), to extract the mean proper lifetime $\tau_{0}$. The corresponding theoretical prediction is derived from Eq.~(2.4), using the scalar decay width calculated according to Eq.~(8.2).

In Figure~16, the lifetime distributions for the signal samples with $m_s = 20~\mathrm{GeV}$ are shown separately for each considered mixing angle. The simulated lifetime distributions are displayed in blue, while the exponential fits are overlaid in red. The extracted values of $\tau_{0}$ from the fits are compared with the theoretical estimates and reported in the legend of each plot.
\\
\begin{figure}[htbp]
	\centering
	\includegraphics[width=0.75\textwidth]{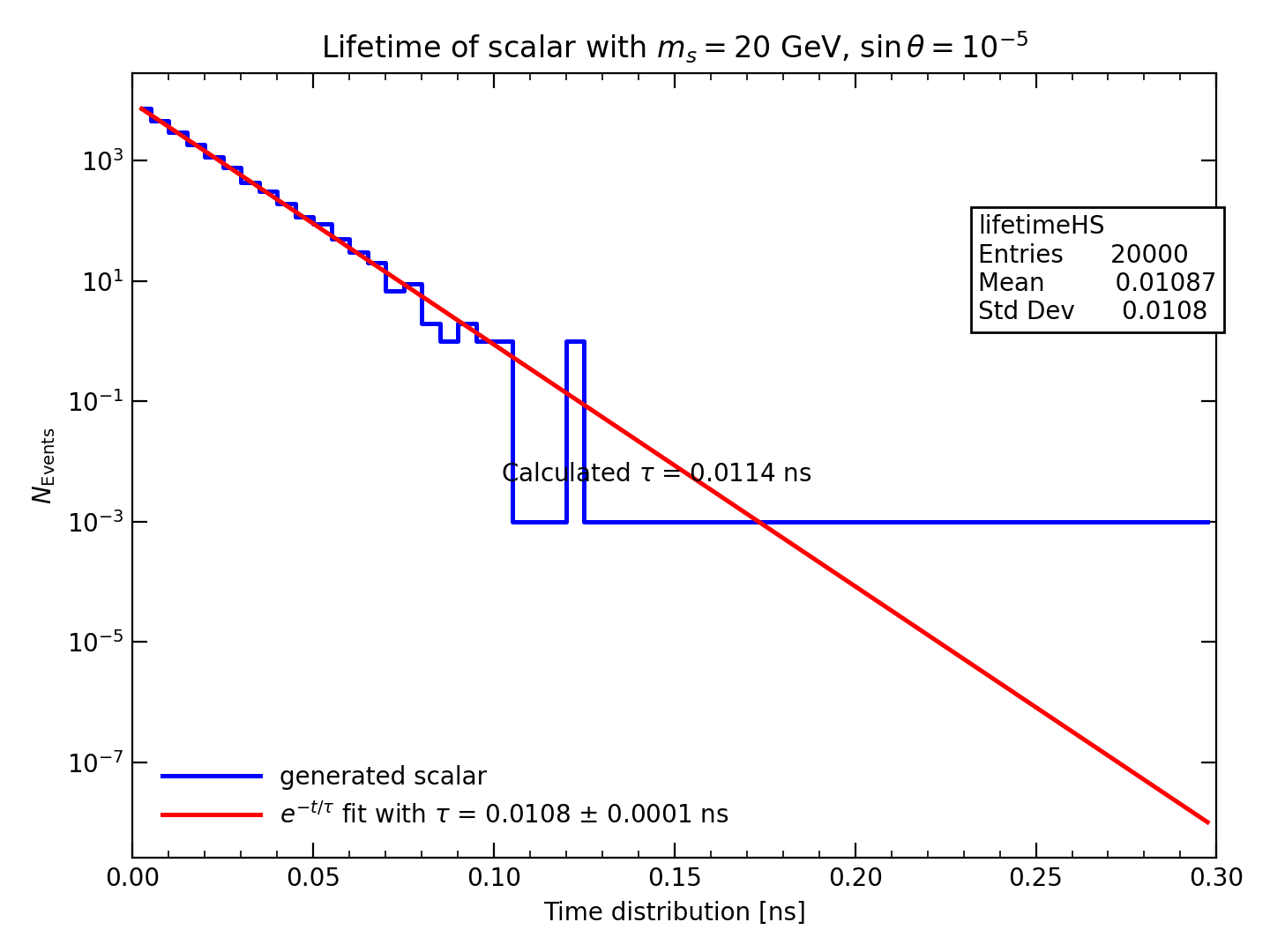}
	\caption{Generated lifetime distribution for the signal sample with $\sin\theta = 1\times10^{-5}$ and
		$m_s = 20~\mathrm{GeV}$.}
	\label{fig:lifetime_ms20_sin1e-5}
\end{figure}
\\
\begin{figure}[htbp]
	\centering
	\includegraphics[width=0.75\textwidth]{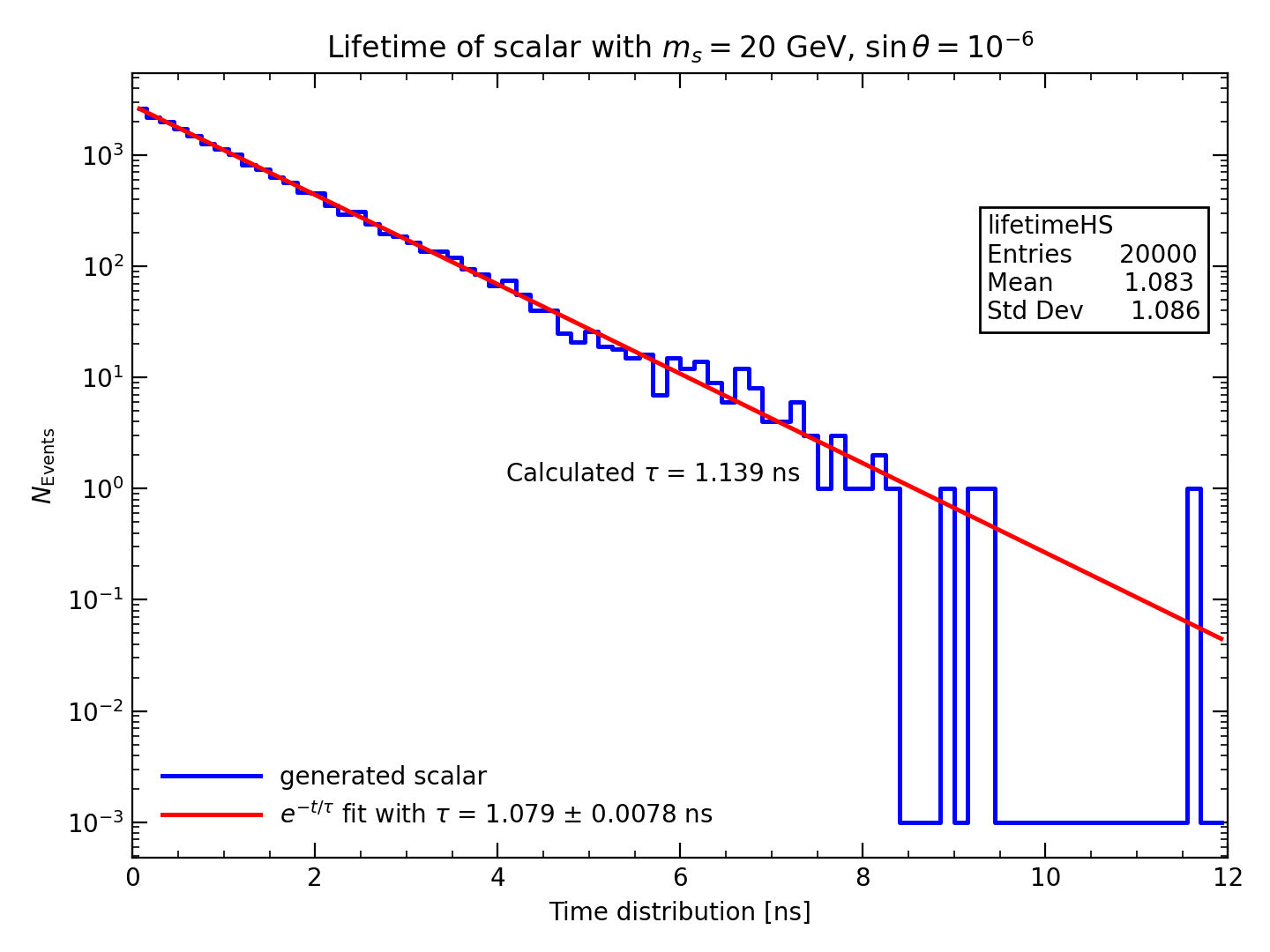}
	\caption{Generated lifetime distribution for the signal sample with $\sin\theta = 1\times10^{-6}$ and
		$m_s = 20~\mathrm{GeV}$.}
	\label{fig:lifetime_ms20_sin1e-6}
\end{figure}
\\
\begin{figure}[htbp]
	\centering
	\includegraphics[width=0.75\textwidth]{lifetime_ms20_sin1e-7}
	\caption{Generated lifetime distribution for the signal sample with $\sin\theta = 1\times10^{-7}$ and
		$m_s = 20~\mathrm{GeV}$.}
	\label{fig:lifetime_ms20_sin1e-7}
\end{figure}
\\
In all three cases, the mean lifetime extracted from the simulation is found to be slightly smaller than the corresponding theoretical prediction, even when statistical uncertainties are taken into account. This discrepancy may arise from the fact that the analytical expressions used to compute the proper lifetime are evaluated at leading order, as given in Eq.~(3.2), and that the branching ratio is taken from Figure~3. In contrast, the event generation includes higher-order effects. Employing more precise calculations for the decay width and branching ratio is therefore expected to improve the level of agreement.

Nevertheless, the relative difference between the theoretical estimates and the fitted values does not exceed 5.2\%. Based on this level of agreement, and in conjunction with the previous validation results, the generated signal samples are considered sufficiently accurate and reliable for the purposes of this analysis.

\section{Conclusion}
In this work, we investigated the process $e^+e^- \to H^+H^-$ within the framework of the scotogenic model. The scattering amplitude receives contributions from tree-level diagrams involving photon and $Z$-boson exchange, as well as from the exchange of the singlet right-handed fermions $N_{1,2,3}$. We examined the experimental and theoretical constraints that significantly restrict the region of parameter space relevant to the process $e^+e^- \to H^+H^-$. In addition, the individual contributions of the photon, the $Z$ boson, and the fermions $N_{1,2,3}$ to the total production cross section were evaluated, taking into account all stringent bounds imposed on the model parameters.

Our analysis shows that the dominant contribution to the cross section arises from the exchange of the singlet right-handed fermions $N_{1,2,3}$. We further studied the dependence of the cross section on the center-of-mass energy for a set of benchmark points that satisfy the most restrictive constraints. Future high-energy $e^+e^-$ colliders will be able to probe the process $e^+e^- \to H^+H^-$ with high precision, thereby providing an opportunity to test these predictions by either strengthening existing bounds or validating the results presented in this study.

\bibliographystyle{apsrev4-2}
\bibliography{su5bib}

\end{document}